# Azimuthal modulation of the event rate of cosmic ray extensive air showers by the geomagnetic field

A.A.Ivanov, V.P.Egorova, V.A.Kolosov, A.D.Krasilnikov, M.I.Pravdin, I.Ye.Sleptsov

Insitute of Cosmophysical Research and Aeronomy, Yakutsk 677891, Russia

a.a.ivanov@sci.yakutia.ru

Abstract

The Earth's magnetic field effect on the azimuthal distribution of extensive air showers (EAS) of cosmic rays has been evaluated using a bulk of the Yakutsk array data. The uniform azimuthal distribution of the EAS event rate is rejected at the significance level $10^{-14}$. Amplitude of the first harmonics of observed distribution depends on zenith angle as $A_1 \approx 0.2 \times sin^2\theta$ and is almost independent of the primary energy; the phase coincides with the magnetic meridian. Basing upon the value of measured effect, the correction factor has been derived for the particle density depending on a geomagnetic parameter of a shower.

The charged particle trajectories of the EAS developing in the atmosphere are distorted in the geomagnetic field. It results in stretched lateral distribution of particles in the plane perpendicular to the shower axis along the direction of the Lorentz force. The observation of this effect at the Yakutsk array in very inclined showers at zenith angles $\theta > 60^o$ has revealed the significant asymmetry [1] of the lateral distribution function (LDF) of particles due to the deviation of muons proportional to the geomagnetic parameter, g, determined by the angle between the shower axis and the field vector: $g = sin\chi / cos^2\theta$, where $\chi = arccos(cos\theta cos\theta_H + sin\theta sin\theta_H cos\alpha)$, $\theta_H=14^0$, $\alpha$ - azimuth.

The geomagnetic effect on the lateral distribution of particles was confirmed analyzing the showers in zenith angle range $20^o<\theta<60^o$ [2]. The calculation results of the expected effect in the QGS model were used to analyze a shower of the highest energy $E_o>10^{20}$ eV observed at the Yakutsk array [3]. In [4 a, b] it was shown that the showers initiated by extremely high energy photons have to show the north-south asymmetry. But the last effect has a threshold for the primary energy: $E_\gamma > 3 \cdot 10^{19}$ eV.

In this paper we draw attention to the fact that due to the LDF asymmetry depending on $\chi$, one can select the showers of equal energy and zenith angle in order to obtain the azimuthal modulation of the EAS event rate because of the different LDF stretch which causes the different observed primary cosmic ray energy and intensity.

Fig.1 shows the distribution of the number of EAS events above $5 \cdot 10^{16}$ eV measured with the Yakutsk array in the period 1974-1995, zenith angles in the bins $20^o$-$30^o$, $40^o$-$50^o$, $60^o$-$70^o$. The number of EAS events in zenith angle intervals used in the analysis is given in Table 1.

Table 1. The number of EAS events (n)

| # | 1 | 2 | 3 | 4 | 5 | 6 | 7 | 8 |
|---|---|---|---|---|---|---|---|---|
| $\theta$, deg. | 0-10 | 10-20 | 20-30 | 30-40 | 40-50 | 50-60 | 60-70 | 70-80 |
| n | 25924 | 69114 | 83226 | 66649 | 37575 | 16384 | 6609 | 1749 |

The amplitude and phase of the first three harmonics of the distribution versus zenith angle are shown in Fig.2, and versus the primary energy – in Fig.3. The second and third harmonics amplitudes come up to the expected value for the uniform distribution of azimuths shown by a dashed line for the measured number of showers in each interval. Dotted lines show the r.m.s. deviations from the expected amplitudes. The first harmonics amplitude essentially differs from zero in zenith angle intervals no. 3 to 7 given in Table 1. In these intervals one can discard the uniform distribution with error probability below $10^{-14}$ basing on the probability for the uniform distribution to have the first harmonics amplitude $A_1$: $P(>A_1)= exp(-n \times A_1^2/4)$. The first harmonics phase coincides with the magnetic meridian in Yakutsk. Almost vertical vector of the magnetic field in the array region $(\theta_H = 14^o)$ results in the dominance of the first harmonics in the range $20^o<\theta<70^o$. For other arrays a situation may be different. For example, at the Tibet array where the field zenith angle is $\theta_H = 45^o$, both the first and the second harmonics at $\theta >50^o$ are equally prevailing. At the Chacaltaya array $(\theta_H =88^o)$ the second harmonics should dominate.

To reveal the influence of the array's geometry on a modulation power, the azimuthal distribution of the event rate was used in a sample of showers registered by detectors arranged in a circle with radius 1.5 km around an array center (azimuth-symmetric part of the array). It turned out that the amplitude and phase of the first harmonics in the sample differ from the corresponding values of the general distribution less than the statistical errors. That is why we've assumed the azimuthal modulation independent of the array geometry.

The data handling of the Yakutsk array used to be carried out fitting the parameters of the axially symmetric function, which approximates the charged particle density in detectors. The particle density at a distance 300 m from the shower core, $\rho_{300}$, found in this way, is connected to the EAS primary energy. Let's consider how $\rho_{300}$ changes because of the geomagnetic scattering of particles with LDF of the form $\rho_r = c \times r^{-\eta}$, where $r$ is a core distance in a perpendicular plane to the EAS axis. The density functions of positively and negatively charged particles are shifted with respect to a shower axis:

$$\rho^g_r = c \times (((r \times cos\varphi + d)^2 + r^2 \times sin^2\varphi)^{-\eta/2} +((r \times cos\varphi-d)^2+r^2 \times sin^2\varphi)^{-\eta/2})/2,$$

where $d$ is a displacement distance of the charges; $\varphi \subset (0^0 \div 360^0)$.
Averaging over a circle of $r = 300$ m we obtain:

$$\rho^g_{300} \approx \rho^{g=0}_{300} \times (1 + (\eta \times d_{300}/2)^2),$$

where $d<<300$; $d_{300}=d/300$ м. On the average, the particle density in the detector increases in comparison with an expected density in the case of absence of the geomagnetic field (i.e. $g=0$). For showers arriving from the north the observed particle densities in detectors are higher than in "southern" showers of the equal energy with the equal zenith angle.

When we select showers with the same observed densities $\rho_{300}$, the primary energy of showers arriving from the north is less. It gives the decrease in a corresponding event rate as it is seen from Fig.1, because of the number of the registered showers at the Yakutsk array diminishes as the energy decreases in the region $10^{16}<E_0<10^{18}$ eV. The amplitude of the first harmonics $A_1$ is defined by a relative value of density change measured for the southern and northern showers: $2 \times A_1 = \Delta J / J = \gamma \times \Delta\rho_{300} / \rho_{300}$, where $J$ is the EAS event rate; $\gamma = lnJ / ln\rho_{300}$. The zenith angle dependence of $A_1$ is well described by $0.2 \times sin^2\theta$ (Fig.2), but the variation of $\gamma$ with zenith angle leads to the ratio $\Delta\rho_{300}/\rho_{300}$, which is not described by the simple dependence.

It is known that the zenith angle boundary, separating showers where electron-photon or muon component dominates, is about $50^0$ at the sea level, $r = 300$ m [5]. Fig.2 shows that the geomagnetic effect in EAS is observed in both these regions. A distance of the charge shift determines the LDF asymmetry caused by the geomagnetic effect. The ratio $\xi$ of the largest distance to the least one corresponding to the fixed particle density $\rho^{g=0}_{300}$ is:
$\xi \approx 1+(1+\eta/2) \times d^2_{300} \approx 1+\Delta\rho_{300}/\rho_{300}$ for $\eta = 3$. So, the measurements of the LDF "oval" can be used to estimate the change of $\rho_{300}$ due to the geomagnetic effect. As it was shown in [1], the linear coefficient between the LDF asymmetry and the geomagnetic parameter is $d\xi/dg= 0.1 \pm 0.04$ for very inclined showers. Using this value for all zenith angles we obtain the correction factor to the density $\rho_{300}$ (Table 2).

Table 2. Correction factors to the measured density $\rho_{300}$

| | $\theta$, deg. | 20 | 30 | 40 | 50 | 60 |
|---|---|---|---|---|---|---|
| $\rho^{g=o}_{300}/\rho^{g}_{300}$ | Northern showers | 0.94 | 0.92 | 0.88 | 0.82 | 0.72 |
| | Southern showers | 0.99 | 0.96 | 0.93 | 0.87 | 0.78 |

As it is seen from Fig.3 the first harmonics amplitude is almost independent of $E_0$ in the energy range where there is sufficient number of events, so the factor dependence on energy can be neglected. The recalculation from $\rho_{300}$ to $E_0$ at the Yakutsk array is carried out using the method of connecting $\rho_{300}$ in inclined and vertical showers through the equi-intensity lines in spectra at the various zenith angles. Because of these intensities correspond to the same energy of the EAS primaries and the geomagnetic corrections can be omitted in vertical showers, then the procedure of $E_0$ estimation correctly takes into account the zenith-angle dependence, but averages the variation of energy with azimuth. It leads to a small systematic error proportional to $A_1$ in estimation of the primary energy. The attenuation length of $\rho_{300}$ obtained by equi-intensity lines is overestimated by ≈10% in comparison to a value for $g$=0.

When the geomagnetic corrections are taken, the EAS size spectrum index for $\rho_{300}$ ($\rho_{600}$ for showers registered with detectors spaced by 1 km distance) increases due to the augmented with $\rho_{300}$ fraction of the inclined showers detected by the array. The resultant corrections to the size spectrum parameters are given in Table 3.

Table 3. The spectrum index ($\gamma$) and intensity (J) corrections

| $E_0$, eV | $10^{18}$ | $3\times10^{18}$ | $10^{19}$ |
|---|---|---|---|
| $\Delta\gamma$ | 0.01 | 0.07 | 0.24 |
| $\Delta J/J$ | 0.16 | 0.19 | 0.28 |

This work is supported in part by the Russian Ministry of Sciences ('Unique Research Facilities in Russia' grant) and RFBR (grant no. 98-07-90315).

**Captions to Figures**

Figure 1. Relative numbers of EAS events in zenith angle intervals (shown at the right) for showers above $5 \cdot 10^{16}$ eV. The function $1+A_1 \times cos(\alpha - \alpha_1)$ is shown by dashed lines.

Figure 2. The amplitudes of the first three harmonics (upper panel) and the phase of the first harmonics (lower panel) versus zenith angle. Statistical errors are shown by the vertical bars, horizontal bars show the angular bins. Symbols are: triangles - the first harmonics; circles – the second harmonics; squares – the third harmonics. The approximation $0.2 \times sin^2\theta$ is shown by the dash-and-dot line. The expected amplitude of harmonics for the uniform distribution in azimuth is given by the dashed line. The standard deviation of expected amplitudes is shown by the dotted lines.

Figure 3. The amplitudes of the first three harmonics and a phase of the first harmonics as a function of energy. The symbols are the same as in Fig.2.

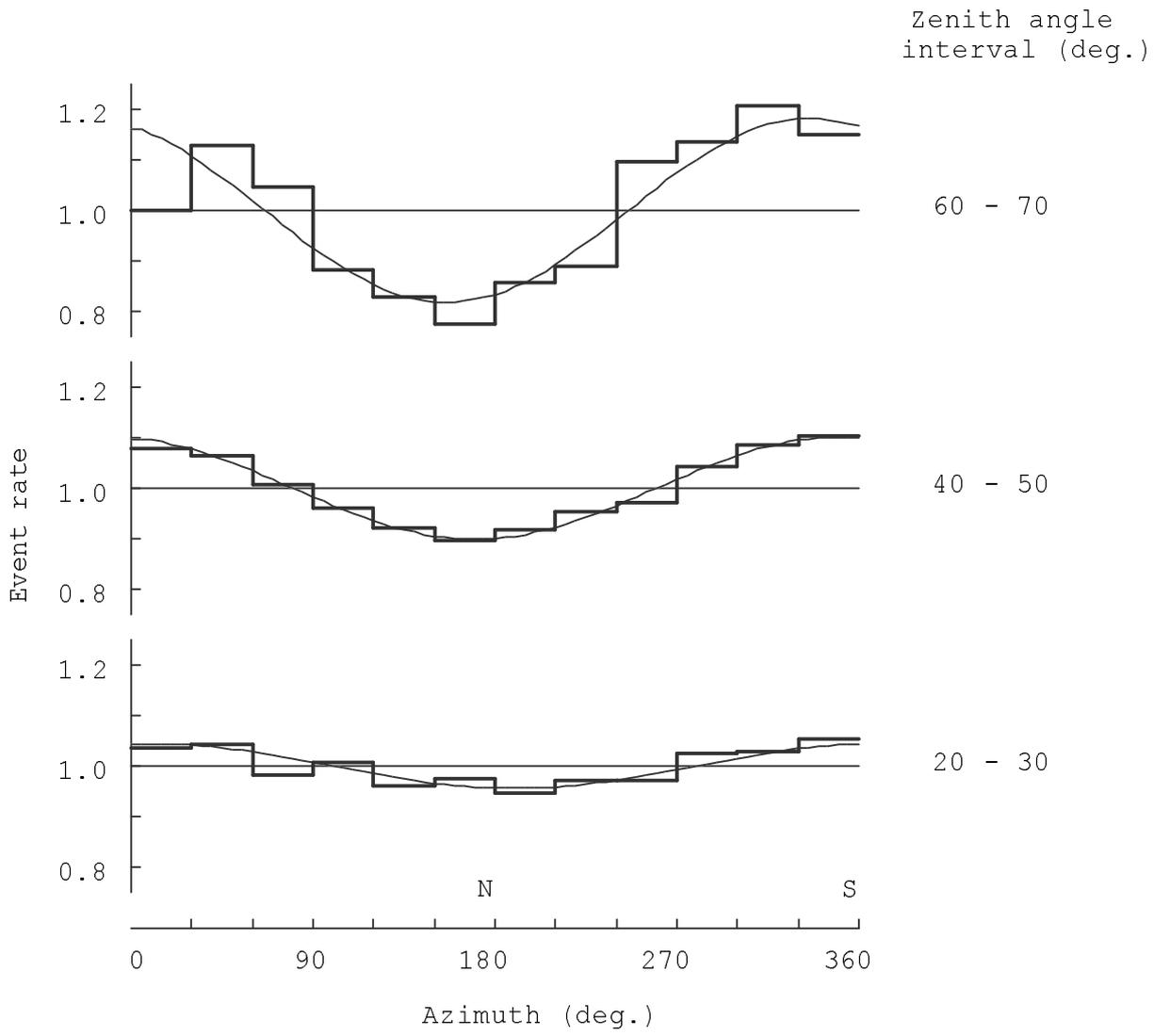

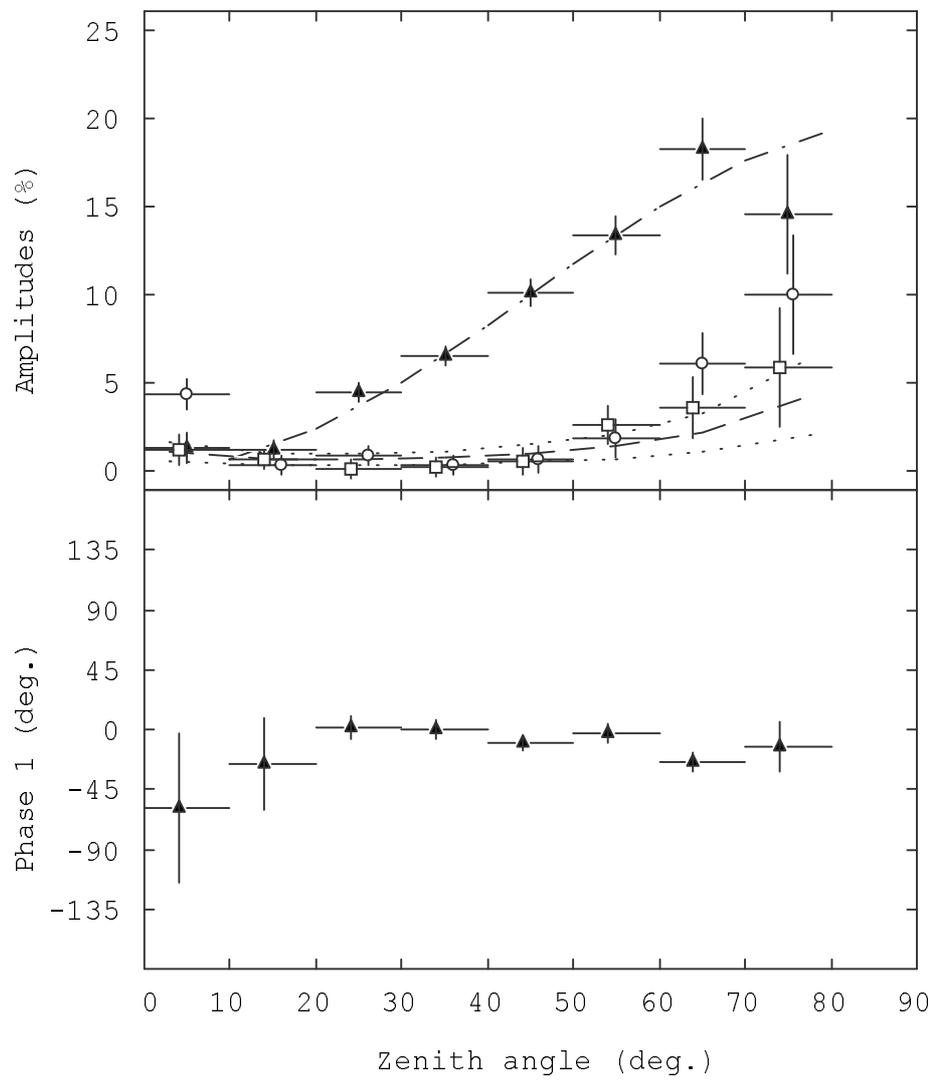

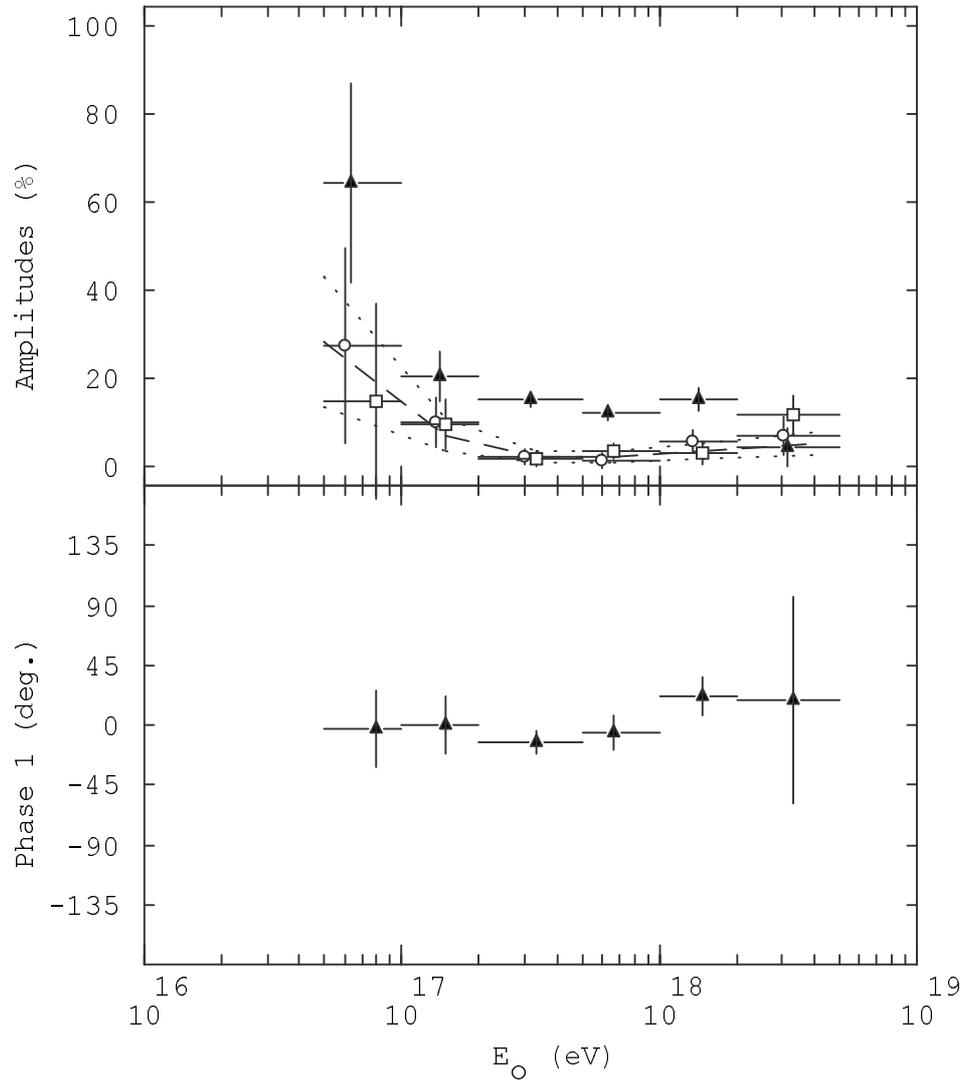